# New Superconductivity in Layered 1$T$-TaS$_{2-x}$Se$_x$ Single Crystals Fabricated by Chemical Vapor Transport**

*Yu Liu, Ran Ang,\* Wenjian Lu, Wenhai Song, Lijun Li, and Yuping Sun\**

**ABSTRACT:** Layered transition-metal dichalcogenides 1$T$-TaS$_{2-x}$Se$_x$ (0 ≤ $x$ ≤ 2) single crystals have been successfully fabricated by using a chemical vapor transport technique in which Ta locates in octahedral coordination with S and Se atoms. This is the first superconducting example by the substitution of S site, which violates an initial rule based on the fact that superconductivity merely emerges in 1$T$-TaS$_2$ by applying the high pressure or substitution of Ta site. We demonstrate the appearance of a series of electronic states in 1$T$-TaS$_{2-x}$Se$_x$ with Se content. Namely, the Mott phase melts into a nearly commensurate charge-density-wave (NCCDW) phase, superconductivity in a wide $x$ range develops within the NCCDW state, and finally commensurate charge-density-wave (CCDW) phase reproduces for heavy Se content. The present results reveal that superconductivity is only characterized by robust Ta 5$d$ band, demonstrating the universal nature in 1$T$-TaS$_2$ systems that superconductivity and NCCDW phase coexist in the real space.

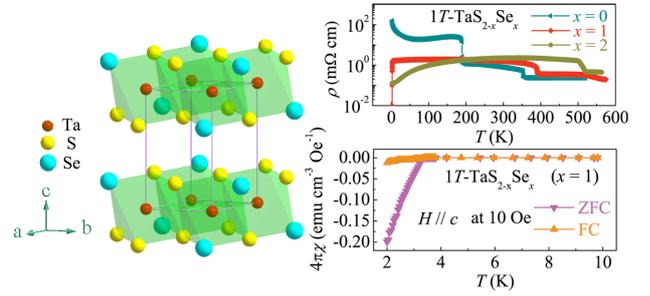

In low-dimensional electron systems, charge-density-wave (CDW) and superconductivity are two of the most fundamental collective quantum phenomena. The CDW is periodic modulation of the density of conduction electrons in solids. The CDW often appears in the proximity of superconductivity near the Mott insulating phase, while the CDW is generally believed to be competing with superconductivity.[1-4] Layered transition-metal dichalcogenide (TMD) provides an excellent platform to investigate the interplay between electron correlation, CDW, superconductivity and other electronic orders, essentially taking into account its quasi-two-dimensional crystal structure susceptible to various electronic instabilities.[5,6]

Amongst many TMD materials, 1$T$-TaS$_2$ and 1$T$-TaSe$_2$ have attracted special attention due to their unique physical properties related to the formation of CDW.[7] Both materials have the same CdI$_2$-type structure with the space group $P$-3$m$1, composed of planes of hexagonally arranged Ta atoms, sandwiched by two chalcogen (S and Se) layers coordinating the central Ta atom in an octahedral arrangement as shown in Figures 1a and 1c. Although these two materials have very similar crystal structure and CDW superstructure, they exhibit dramatically different physical properties. In particular, 1$T$-TaS$_2$ is characterized by a high-temperature normal metallic phase, followed by a nearly commensurate CDW (NCCDW) phase and commensurate CDW (CCDW) phase coexisting with a Mott insulating phase. In comparison, 1$T$-TaSe$_2$ undergoes a high-temperature normal metal to CCDW transition without drastic change in the electronic conductivity in the whole CCDW phase, and presents a metallic behavior.

The Mott insulating phase of 1$T$-TaS$_2$ is unusual, Ta atoms are displaced to form David-star clusters where twelve Ta atoms contract toward central atom, and each clusters are interlocked by forming triangular superlattice with a √13 × √13 periodicity,[7-11] accompanied by the narrowing of the Ta 5$d$ valence band and the formation of multiple subbands due to band folding.[12,13] With increasing the temperature, the Mott phase of 1$T$-TaS$_2$ melts into the NCCDW phase with a sudden resistivity drop, where several tens of stars organize into roughly hexagonal domains.[14] On the other hand, a strong CCDW phase of 1$T$-TaSe$_2$ arises until high temperature, holding a √13 × √13 superstructure,[7,15-17] analogous to the low-temperature CCDW phase of 1$T$-TaS$_2$.

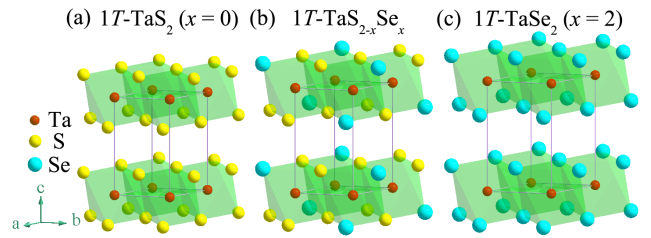

**Figure 1.** Crystal structure of (a) pristine 1$T$-TaS$_2$ ($x$ = 0), (b) 1$T$-TaS$_{2-x}$Se$_x$, and (c) pristine 1$T$-TaSe$_2$ ($x$ = 2).

[*] Dr. Y. Liu, Dr. R. Ang, Dr. W. J. Lu, Dr. W. H. Song, Dr. L. J. Li, Prof. Y. P. Sun
Key Laboratory of Materials Physics
Institution of Solid State Physics, Chinese Academy of Sciences，Hefei 230031, People's Republic of China
Fax: (+86 551 559 2757)
E-mail: rang@issp.ac.cn; ypsun@issp.ac.cn.

Prof. Y. P. Sun
High Magnetic Field Laboratory, Chinese Academy of Sciences，Hefei 230031, People's Republic of China

[**] This work was supported by the National Key Basic Research under contract No. 2011CBA00111, and the Joint Funds of the National Natural Science Foundation of China and the Chinese Academy of Sciences' Large-Scale Scientific Facility (Grant No. U1232139).

Supporting information for this article is available.

Until now, superconductivity emerges in 1$T$-TaS$_2$ by applying high pressure[18] or introducing disorders in the crystal.[19] It has been recently reported that the substitution of Ta with extremely light Fe content in 1$T$-Fe$_x$Ta$_{1-x}$S$_2$ (0 ≤ $x$ ≤ 0.05) induces superconductivity in the NCCDW state on the verge of Mott and Anderson localization.[20] More importantly, the high-resolution angle-resolved photoemission spectroscopy (ARPES)[21] directly accessed the essential issue to elucidate the electronic states of all the available electronic phases in 1$T$-Fe$_x$Ta$_{1-x}$S$_2$, where the superconductivity is characterized by a shallow electron pocket at the Brillouin-zone center, the superconductivity and the NCCDW state coexist in the real space. Even so, an investigation on the emergence of superconductivity by the substitution of S site has not yet been made. We wondered the superconductivity in 1$T$-TaS$_2$ whether it is the intrinsic universality and not dependent on the substitution of Ta site or S site.

In this work, we focused on the substitution of S site in a whole Se range. We succeeded in growing a series of new 1$T$-TaS$_{2-x}$Se$_x$ ($0 \leq x \leq 2$) single crystals by using a chemical vapor transport (CVT) technique. We discovered the superconducting evidence for the first time by the substitution of S site. Moreover, we observed the various electronic states with Se content and summarized the rich phase diagram. The transport and magnetic measurements reveal that the essential superconductivity in 1$T$-TaS$_2$ systems is only related to Ta 5$d$ band, regardless of the substitution of Ta site or S site. These results further demonstrate the universal feature in 1$T$-TaS$_2$ systems that superconductivity and NCCDW phase coexist in the real space.

The crystal structure of 1$T$-TaS$_{2-x}$Se$_x$ resembles the pristine 1$T$-TaS$_2$ and 1$T$-TaSe$_2$ (see Figure 1b). As shown in Figure S1a in the Supporting Information, only (00l) reflections were observed in the single-crystal XRD patterns for each sample, indicating that the crystallographic $c$ axis is perpendicular to the plane of the single crystal. With increasing the Se content, the diffraction peak along the (00l) direction distinctly shifts to lower angle (see Figure S1b), confirming the larger lattice parameter of Se than that of S.

Structural refinement of powder XRD identifies that all reflections can be indexed in the $P$-3$m1$ space group. Figures S2a-S2c show the XRD patterns and the structural refinement results of Rietveld analysis for the selected samples with $x$ = 2, 1, and 0. For the Rietveld fitting, Ta and S (Se) atoms are placed at the 1$a$ (0, 0, 0) and 2$d$ (1/3, 2/3, 1/4) positions, respectively. The refined structural parameters and selected bond lengths and angles are listed in Table S1. For $x$ = 0, the lattice parameters are $a$ = 3.3672(6) Å and $c$ = 5.9020(9) Å. For $x$ = 1, the lattice parameters are $a$ = 3.4173(8) Å and $c$ = 6.1466(4) Å. For $x$ = 2, the lattice parameters are $a$ = 3.4746(9) Å and $c$ = 6.2778(4) Å. Comparatively, the lattice distortion along $c$ axis is stronger than that of $ab$ plane from $x$ = 0 to $x$ = 2. In Table S1, it can reflect the information of lattice distortion by the decrease of bond angles ($\theta$) from $\theta_{S-Ta-S}$ (87.234°) to $\theta_{Se-Ta-Se}$ (86.014°). Figure S2d shows the variation of lattice parameters ($a$, $c$) and unit cell volume ($V$) with Se content. The values of $a$, $c$, and $V$ monotonously increase with Se content, in accordance with the larger ion radius of Se than that of S. In Figure 1, it is also clearly seen the evolution of crystal structure along $c$ axis and $ab$ plane from 1$T$-TaS$_2$ to 1$T$-TaSe$_2$. This is totally different from the case of the substitution of Fe at Ta site in 1$T$-Fe$_x$Ta$_{1-x}$S$_2$,[20] where the values of $c$ almost have no change with increasing Fe content, while the values of $a$ decrease due to the smaller ion radius of Fe than that of Ta.

The temperature dependence of in-plane resistivity ($\rho_{ab}$) measurements on 1$T$-TaS$_{2-x}$Se$_x$ ($0 \leq x \leq 2$) is shown in Figure 2a. At first below 200 K, it is observed a first-order transition from the NCCDW to the CCDW phase. The CCDW transition gradually shifts to lower temperature with increasing $x$ content, and the Mott insulating phase at low temperatures completely disappears for $x$ > 0.8. More significantly, the signature of superconductivity emerges for $x$ = 0.9. The maximum of superconducting onset temperature is ~ 3.6 K for the optimal sample with $x$ = 1 (see the inset of Figure 2a). Above $x$ > 1.6, the superconductivity at low temperatures is suppressed, while CCDW phase reproduces and a metallic behavior arises until $x$ = 2. Figure 2b shows the high-temperature resistivity to magnify the NCCDW transition from normal metallic phase. The NCCDW transition obviously shifts to higher temperature with Se content, in agreement with the larger ion radius of Se than that of S.

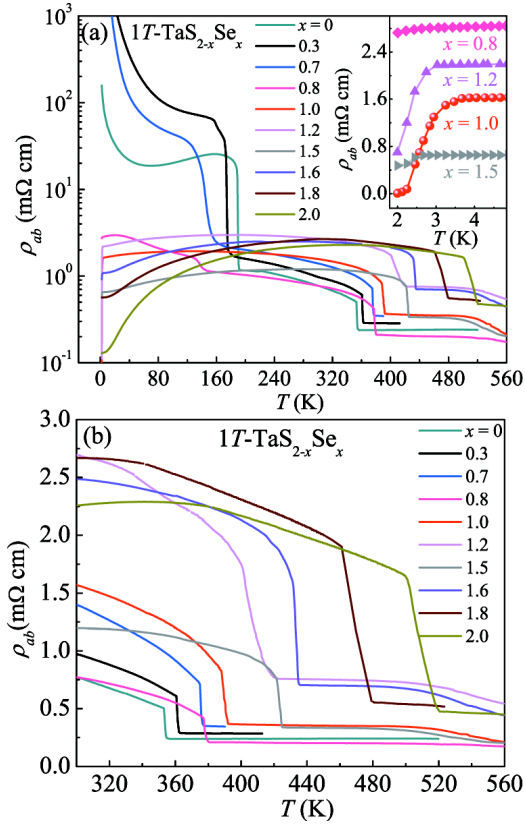

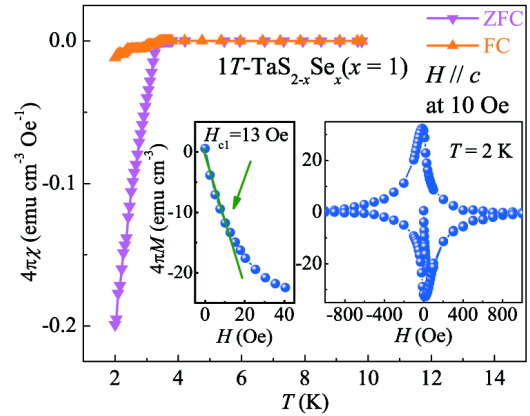

*Figure 2.* (a) Temperature dependence of in-plane resistivity ($\rho_{ab}$) of 1$T$-TaS$_{2-x}$Se$_x$. The inset magnifies the region of the superconducting transitions. (b) The nearly commensurate CDW (NCCDW) transitions at high temperatures of all the samples.

*Figure 3.* Temperature dependence of magnetic susceptibility for optimally superconducting sample 1$T$-TaSSe ($x$ = 1). The left inset shows the initial $M(H)$ isotherm at 2 K, where the light green line stands for the linear fitting in the low-field range. The right inset shows the magnetization hysteresis loops of 1$T$-TaSSe at 2 K with $H$ parallel to the $c$-axis.

Figure 3 shows the magnetic properties of optimally superconducting sample for 1$T$-TaSSe ($x$ = 1) at 10 Oe with magnetic field $H$ parallel to the $c$-axis. Undoubtedly, the diamagnetism signals at low temperatures demonstrate the occurrence of superconductivity, corresponding with the resistivity data. The steep transition reveals that the sample is rather homogeneous. The superconducting transition temperature $T_c$ defined by the onset point of ZFC and FC curves is ~ 3.5 K for 1$T$-TaSSe ($x$ = 1), which is higher than that of $T_c$ (~ 2.1 K) for the optimally superconducting 1$T$-Fe$_{0.02}$Ta$_{0.98}$S$_2$.[20] The left inset of Figure 3 shows the initial $M(H)$ curve of $x$ = 1 in the low-field region at 2 K. We can get the value of lower critical field ($H_{c1}$ ~ 13 Oe), marked by an arrow from the point where this curve deviates from linearity. The obtained slope of the linear fitting up to 13 Oe for present experimental data is -0.955, which is accordant with -4$\pi M$ = $H$, describing the Meissner shielding effect. The right inset of Figure 3 shows the magnetization hysteresis loop for $x$ = 1 at 2 K. The shape of $M(H)$ curve further demonstrates the present 1$T$-TaSSe sample is a typical type-II superconductor.

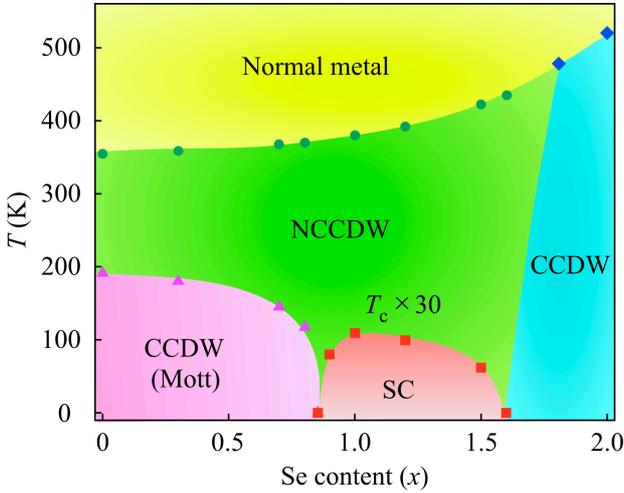

**Figure 4.** Electronic phase diagram of 1$T$-TaS$_{2-x}$Se$_x$ derived from the present transport properties as a function of temperature and $x$, where CCDW, NCCDW, and SC represents the commensurate CDW, nearly commensurate CDW, and superconductivity, respectively.

Figure 4 summarizes the overall electronic phase diagram as a function of temperature and $x$ in 1$T$-TaS$_{2-x}$Se$_x$. The Mott localization and the CCDW phase are evidently suppressed for $x$ > 0.8. After that, a new superconducting state emerges in a wide range until for $x$ = 1.6. The $x$ dependence of superconducting transition temperature $T_c$ follows a dome-like shape. Ultimately, the superconductivity disappears and CCDW phase is induced again. In fact, the $T_c$-dome shape is also observed in the superconducting 1$T$-Fe$_x$Ta$_{1-x}$S$_2$ by the substitution of Ta site, although it only occurs in a very narrow $x$ range from $x$ = 0.01 to $x$ = 0.04. It is demonstrated that the extra carriers induced by Fe substitution are not responsible for the emergence of superconductivity. In particular, no any experimental evidence on Fe 3$d$ band is observed near the Fermi level, and only Ta 5$d$ band contributes to the superconductivity.[21] That is to say, the superconductivity is not dependent on the substitution of S or Ta site. Therefore, it is very important to clarify the electronic structure of universal superconducting 1$T$-TaS$_2$ systems. We will discuss implications of superconductivity later.

In the superconducting region of 1$T$-Fe$_x$Ta$_{1-x}$S$_2$, it is found a single electron pocket at the Brillouin-zone center Γ point in the NCCDW phase (melted Mott phase) by the ARPES measurement.[21] This electron pocket is not observed in the normal metallic phase and Mott phase, it is characteristic of the NCCDW phase, and likely created by the backfolding of bands due to the superlattice potential of NCCDW. Furthermore, the strong ARPES intensity is only observed around Γ point in the NCCDW phase.[21] This suggests that the shallow electron pocket at the Γ point dominates the transport properties in the NCCDW phase. Naturally, it is to conclude that the superconductivity is also characterized by the shallow electron pocket at the Γ point, since the superconducting region appears inside the NCCDW phase.

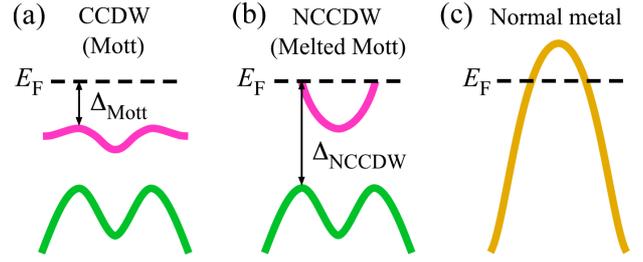

**Figure 5.** Schematic band diagram of 1$T$-TaS$_{2-x}$Se$_x$ in (a) CCDW (Mott) phase, (b) NCCDW (Melted Mott) phase, and (c) normal metallic phase, respectively. The $\Delta_{Mott}$ and $\Delta_{NCCDW}$ denotes the Mott gap and NCCDW gap, respectively.

Taking into account the universal superconducting nature for the substitution of S or Ta site in 1$T$-TaS$_2$ systems, the schematic band diagram of 1$T$-TaS$_{2-x}$Se$_x$ is shown in Figure 5. In the normal metallic phase, it is a weakly correlated metal where the one-electron approach entirely explains the band dispersion (see Figure 5c). In the NCCDW (melted Mott) phase, a shallow electron pocket appears and characterizes the transport properties of the NCCDW phase as well as superconductivity as shown in Figure 5b. The electron pocket is gapped in the Mott phase (see Figure 5a) due to the enhanced electron correlation, implying the competing nature of superconductivity with Mott localization.

Finally, we discuss the implications of the shallow electron pocket in the NCCDW phase. The superconductivity should be characterized by the electron pocket by considering (i) its dominant intensity contribution in the whole momentum space in the NCCDW phase and (ii) the $x$ dependence of the electron-pocket density of states following the superconducting-dome shape.[21] In the real space, the NCCDW and superconductivity coexist within the star-of-David clusters. The superconductivity of 1$T$-TaS$_{2-x}$Se$_x$ and 1$T$-Fe$_x$Ta$_{1-x}$S$_2$ is only reflected by the robust Ta 5$d$ band, while it is not related to the S 4$p$ and Fe 3$d$ band. Moreover, the shallow electron pocket transforms into the lower Hubbard band in the Mott phase further reveals the intimate relationship between electron correlations and the superconductivity in 1$T$-TaS$_2$ system (see Figures 5a and 5b). From this viewpoint, the ARPES experiment in 1$T$-TaS$_{2-x}$Se$_x$ should be expected in future.

In summary, new layered 1$T$-TaS$_{2-x}$Se$_x$ (0 ≤ $x$ ≤ 2) single crystals have been successfully prepared by using the CVT technique. This is the first time to report the superconductivity by the substitution of S site. Different electronic states exist in 1$T$-TaS$_{2-x}$Se$_x$: the Mott phase melts into the NCCDW phase, superconductivity in a wide $x$ range develops within the NCCDW phase, and finally CCDW phase reproduces for heavy Se content. The present important results reveal that the superconductivity is not dependent on the substitution of S or Ta site. The universal superconductivity of 1$T$-TaS$_2$ is only characterized by robust Ta 5$d$ band, which is not determined by the band of substituted element such as Se 4$p$ or Fe 3$d$. The universal nature in 1$T$-TaS$_2$ systems further demonstrates that superconductivity and NCCDW phase coexist in the real space. The present 1$T$-TaS$_{2-x}$Se$_x$ provides a new insight into the interplay between electron correlation, CDW, and superconductivity.

## Experimental Section

High-quality single crystals of 1$T$-TaS$_{2-x}$Se$_x$ (0 ≤ $x$ ≤ 2) were grown by the CVT method with iodine as a transport agent. Stoichiometric amounts of the raw materials, high-purity elements Ta, S and Se, were mixed and heated at 900 °C for 4 days in an evacuated quartz tube. Then the obtained TaS$_{2-x}$Se$_x$ powders and iodine (density: 5 mg/cm$^3$) were sealed in another longer quartz tube, and heated for 10 days in a two-zone furnace, where the temperature of source zone and growth zone was fixed at 950 °C and 850 °C, respectively. A shiny mirror-like sample surface was obtained, confirming their high quality.

The crystal structure of 1$T$-TaS$_{2-x}$Se$_x$ single crystals was identified by x-ray diffraction (XRD). The XRD patterns were obtained on a Philips X'pert PRO diffractometer with Cu $K_\alpha$ radiation ($\lambda$ = 1.5418 Å) at room temperature. Phase identity and purity were examined by powder XRD. Structural refinements of powder 1$T$-TaS$_{2-x}$Se$_x$ samples was performed by using Rietveld method with the X'Pert High Score Plus software. The average stoichiometry was determined by examination of multiple points using x-ray energy dispersive spectroscopy (EDS) with a scanning electron microscopy (SEM). The EDS results indicate that the actual concentration $x$ is close to the nominal one. The resistivity measurements ($\rho$) down to 2.0 K were carried out by the standard four-probe method in a Quantum Design Physical Property Measurement System (PPMS). The magnetic susceptibility ($\chi$) was measured in both zero-field-cooled (ZFC) and field-cooled (FC) modes down to 2.0 K using a Quantum Design superconducting quantum interference device (SQUID) system with applied magnetic field ($H$) parallel to the $c$-axis of sample, as well as the hysteresis loop.